\documentclass[12pt]{iopart}
\usepackage{iopams}  
\usepackage{graphicx}  
\usepackage{amsfonts}  
\usepackage{mathrsfs}  

\newcommand{\ps}{\ps} 
\newcommand{\dk}{$\delta$-kicked rotor} 
 
\newcommand{\Pop}{{\hat{\rm P}}}

\newcommand{\Hop}{{\hat H}}

\begin{document} 
 
\title[Can quantum fractal fluctuations be observed \dots ?]{Can quantum fractal fluctuations be observed in an atom-optics kicked rotor experiment?}

\author{Andrea Tomadin\dag{}\ddag{}, Riccardo Mannella\ddag{}, 
and Sandro Wimberger\ddag{}\footnote[3]{Corresponding author's e-mail: saw@df.unipi.it}} 
 
\address{\dag{}Scuola Normale Superiore, Piazza dei Cavalieri, I-56100 Pisa}
\address{\ddag{}CNR-INFM and Dipartimento di Fisica ``Enrico Fermi'', 
Universit\`a degli Studi di Pisa, Largo Pontecorvo 3, I-56127 Pisa}
 
\begin{abstract}
  We investigate the parametric fluctuations in the quantum survival probability of an open version of the \dk{} model in the deep quantum regime.
  Spectral arguments [Guarneri I and Terraneo M 2001 {\it Phys. Rev. E} \textbf{65} 015203(R)] predict the existence of parametric fractal fluctuations owing to the strong dynamical localisation of the eigenstates of the kicked rotor.
  We discuss the possibility of observing such dynamically-induced fractality in the quantum survival probability as a function of the kicking period for the atom-optics realisation of the kicked rotor. 
  The influence of the atoms' initial momentum distribution is studied as well as the dependence of the expected fractal dimension on finite-size effects of the experiment, such as finite detection windows and short measurement times. 
  Our results show that clear signatures of fractality could be observed in experiments with cold atoms subjected to periodically flashed optical lattices, which offer an excellent control on interaction times and the initial atomic ensemble.
\end{abstract} 

 
 
\section{Introduction} 
\label{sec:1}

Experiments with cold atoms nowadays offer unique possibilities for the study of single particle motion and collective particle dynamics in tailored optical or magnetic potentials.
The atomic centre-of-mass motion can be prepared and controlled with unprecedented precision, what allows experimentalists to realise and study many toy models of condensed matter physics \cite{smodels}.
Since in experiments with cold atomic gases noise and perturbations can be driven to a minimum, which often is indeed negligible, such setups offer a great advantage with respect to solid-state realisations.

In this paper, we discuss the possibility of observing sensitive quantum effects which manifest in a fractal variation of a transport function with respect to a well-tunable control parameter.
Similar fractal fluctuations of the transmission probability across solid-state samples have been measured recently \cite{HHHJ1996} in systems whose underlying classical phase space typically contains mixed regular-chaotic structures. 
Most features of these experiments can be understood semiclassically as a consequence of the phase space topology \cite{SKGF1998,K1996,HKW2001}. 
However, the precise origin of the observed fractal conductance fluctuations in these experiments is not yet fully understood \cite{MTMN2004}, and, in fact, various theoretical models \cite{GT2001,BN2003} predict fractal conductance fluctuations for mesoscopic devices. 
Our aim is to design a concrete experimental scenario in which parametric fractal fluctuations could be measured with high precision cold-atom setups. 
In such experiments the cross-over between mixed and completely chaotic classical dynamics can be scanned easily \cite{KOSR1999,SWPL2005,BRMS1999}, and hence fractal transmission probabilities could be measured in a regime where classical or semiclassical arguments do not apply.

As was shown by Guarneri and Terraneo \cite{GT2001}, fractal fluctuations in the transmission probability of a quantum scattering problem arise naturally as a consequence of the spectral properties of the system.
The two essential conditions on the spectrum are (i) a power-law distribution of decay widths and (ii) uncorrelated real parts of the energy spectrum.  
Moreover, various eigenstates have to contribute together to the decay, a fact which is expressed formally by requiring that (iii) the average decay width is much larger than the mean level spacing. 
Based on these conditions, the theory of \cite{GT2001} explained the occurrence of quantum fractal fluctuations in the \dk{} model in the deep quantum realm \cite{BCGT2001}, where semiclassical arguments cannot explain the occurrence of fractality.

In this paper we study a similar dynamical situation as in \cite{BCGT2001}, yet with important modifications which fully account for the actual experimental realisation of the kicked rotor. 
Using either cold or ultracold atomic gases, the kicked rotor is realised by preparing a cloud of atoms with a small spread of initial momenta, which is then subjected to a one-dimensional optical lattice potential, flashed periodically in time \cite{MRBS1995}.
Let us call $k_{L}$ the wave number of the optical lattice, $\tilde{\tau}$ the flashing period (``kicking'' period), $\tilde{p}$ the momentum of the single atom, $\tilde{x}$ its centre-of-mass position, $V_{0}$ the maximum potential depth, and $M$ is the atomic mass. 
It is convenient to adopt rescaled units by noting that $p_{R}=\hbar k_{L}$ is the photon recoil momentum and $E_{R}=(\hbar k_{L})^{2}\,/\,2M$ is the recoil energy \cite{KOSR1999,BRMS1999,AGOS2001}.
So we define $p=\tilde{p}\,/\,2p_{R}$, $x=\tilde{x}\,/\,2k_{L}$, $\tau=\tilde{\tau}\cdot 8E_{R}\,/\,\hbar$. 
The kicking strength of the lattice is expressed by $k=V_{0}\tau\,/\,\hbar$.
The Hamiltonian now reads in dimensionless units \cite{GSZ1992}
\begin{eqnarray}
\label{eq:hamiltonian}
\Hop(t^{\prime}) 
= \frac{p^2}{2} + k\,\cos{x}\;\sum_{t=1}^{\infty}\delta(t^{\prime}
 - t\,\tau)\;.
\end{eqnarray}
Owing to the $\delta$-interaction of the potential with the atoms, the time evolution operator between kicks can be explicitly written in a factorised form, extremely convenient for numeric simulations. The derivation of the one-period evolution operator exploits the spatial periodicity of the potential by Bloch's theorem \cite{BRMS1999,WGF2003}.
This defines \emph{quasimomentum} $\beta$ as a constant of the motion, the value of which is the fractional part of the physical momentum $p$ in dimensionless units 
$p=n+\beta \;\;\; (n \in \mathbb{N})$.
Since $\beta$ is a conserved quantum number, $p$ can be labelled using its integer part $n$ only.
The spatial coordinate is then substituted by $\theta=x\;\mbox{mod}({2\pi})$ and the momentum operator by 
$\hat{\mathcal{N}}=-i\partial/\,\partial\theta$ with periodic boundary conditions. The one-kick propagation operator for a fixed quasimomentum $\beta$ is thus given by \cite{WGF2003}
\begin{eqnarray}
\label{eq:evol}
\hat{\mathcal{U}}_{\beta} = \e^{-ik\,\cos(\hat{\theta})}\;e^{-i\tau (\hat{\mathcal{N}}+\beta)^{2}/\;2} \;.
\end{eqnarray}

In close analogy to the transport problem across a solid-state sample, we follow \cite{BCGT2001} to define the quantum survival probability as the fraction of the atomic ensemble which stays within a specified region of momenta while applying absorbing boundary conditions at the ``sample'' edges.
If we call $\psi(n)$ the wave function in momentum space and $n_{1}<n_{2}$ the edges of the system, absorbing boundary conditions are implemented by the prescription $\psi(n)\equiv 0$ if $n\leq n_{1}$ or $n\geq n_{2}$. 
This truncation is carried out after \emph{each} kick.
This procedure mimics the escape of atoms out of the spatial region where the dynamics induced by the Hamiltonian (\ref{eq:hamiltonian}) takes place.
If we denote by $\Pop$ the projection operator on the interval $]n_{1},n_{2}[$ the survival probability after $t$ kicks is:
\begin{eqnarray}
P_{\rm surv}(\tau;t)= \left \| (\Pop \hat{\mathcal{U}}_{\beta}) ^{t} \psi(n,0;\tau) \right \|^{2} \;.
\end{eqnarray}

We will show in the following that signatures of fractality in the survival probability could be observed in modern atom-optical experiments, where the initial atomic ensemble has a finite, non-zero width in momentum space. 
In contrast to the work of \cite{BCGT2001}, where the initial quasimomentum is scanned to arrive at the parametric observable $P_{\rm surv}(\beta)$, we investigate the behaviour of $P_{\rm surv}(\tau)$ as a function of the best controllable parameter in the experiment, namely the time $\tau$ which elapses between two successive kicks \cite{SWPL2005,AGOS2001,OSR2003,RAVA2005}.

After a brief review of the results of Guarneri and Terraneo \cite{GT2001} applied to the dynamically localised kicked rotor (section \ref{sec:2}), we discuss in section \ref{sec:3} our choices of the system parameters, which are guided by the experimental possibilities as well as the conditions stated in \cite{GT2001}.
Our central results on the occurrence of fractal survival probabilities are presented for the limit of long-interaction times (section \ref{sec:4}) as well as for experimentally accessible initial momentum distribution and interaction times (section \ref{sec:5}). 
Section \ref{sec:6} finally concludes the paper.

\section{Conditions for fractal fluctuations of the survival probability}
\label{sec:2}

Without a priori assumptions on the integrability or chaoticity properties of the classical analogue of the quantum system of interest, Guarneri and Terraneo \cite{GT2001} showed that fractal conductance fluctuations occur if certain conditions on the quantum spectrum of the open system are fulfilled.

The first condition, (i) a power-law distribution of the decay widths, is indeed present in the weakly opened quantum kicked rotor \cite{CMS1999}.
We verified this by diagonalising the one-kick evolution operator $\hat{\mathcal{U}}_{\beta}$, after representing it in the basis of momentum states.
The matrix was cut at the positions $n_{1}$ and $n_{2}$ to mimic the required absorbing boundary conditions. 

If either of the two cutoffs ($n_{1}$ or $n_{2}$) is chosen sufficiently large, the shape of the wavefunction in momentum space supports an exponential tail, independent of the evolution time (after a short transit time $\propto k^{2}$ at which dynamical localisation has fully developed \cite{I1990,F1993}).
For such a situation in the localised regime, the probability density of decay widths was found to be $\rho (\Gamma ) \propto \Gamma^{-1}$ over more than 10 orders of magnitude in $\Gamma$, consistent with previous studies \cite{HKW2001,CMS1999,T2001,TF2000,WKB2002}.
If, on the other hand, $n_{1}$ and $n_{2}$ were decreased, dynamical localisation is gradually destroyed and the distribution deforms continuously, giving more weight to larger widths and less to the very small ones.
Such a deformation was observed in the analogous context of ionisation rates of microwave-driven hydrogen Rydberg atoms \cite{WKB2002}.
Our choice of $n_{1}$ and $n_{2}$ represents a compromise between the maximum width of typical experimental detection windows in momentum space and a guaranteed dynamically localised momentum distribution over a substantial interval of momenta.
In the next section, we state the precise values of $n_{1}$ and $n_{2}$ which we investigated in this paper.

In the regime of strong dynamical localisation, the quasienergy spectrum of the \dk{} has a Poisson-like statistics \cite{FFGP1985}.
Under the same conditions as stated above on the cutoff values $n_{1}$ and $n_{2}$, this property of the real-parts of the quasienergy spectrum remains even when the system is opened \cite{T2001,GTWunp}.
Hence, also the second requirement for fractality of \cite{GT2001}, that (ii) the energy spectrum consists of uncorrelated sequences, is fulfilled in good approximation for the opened \dk{} in the presence of dynamical localisation.

The third condition stated in \cite{GT2001} is that the opening of the system is weak, but still sufficient to guarantee that (iii) the average decay width 
is much larger than the mean level spacing.
For our choice of parameters and cutoff values $n_{1}$ and $n_{2}$, also this condition of overlapping ``resonance peaks'' is fulfilled, as we verified numerically from the quasienergy spectrum of the truncated matrix representation of $\hat{\mathcal{U}}_{\beta}$.

As exercised in \cite{GT2001}, the conditions (i-iii) are sufficient to guarantee self-affine fluctuations in the quantum survival probability, with a predicted fractal dimension $D_{\rm f}$ which is related to the exponent of the width distribution $\rho(\Gamma)\propto\Gamma^{-\alpha}$ by the following general formula $D_{\rm f} = 1+\alpha\,/\,2 \approx 1.5$ for $\alpha \approx 1$.

We repeat that parametric fractal fluctuations in the survival probability of dynamically localised kicked rotor have already been found in \cite{BCGT2001}, before their origin could be explained in \cite{GT2001}.
In this work, however, we scan a different parameter than the one used in \cite{BCGT2001}, which corresponded to quasimomentum. 
Here we use the kicking period $\tau$ as control parameter, which can be much better controlled in state-of-the-art experiments \cite{SWPL2005, AGOS2001,OSR2003,RAVA2005} than the initial value of momentum \cite{BRMS1999,RAVA2005,AGSG2004,WSPL2005,DMCW2004}.
On the other hand, the use of $\tau$ confronts us with a new problem which is discussed in the following section.

\section{Choice of parameters} 
\label{sec:3}

\subsection{Dynamical localisation and classical chaos}
\label{sec:3.1}

For our analysis the value of the kicking strength $k$ was chosen in the range $2\dots 6$, or $k\,\tau= 2.8\dots 8.4$, going along with the transition from local to global chaos with increasing $k$ in this range \cite{I1990,F1993}.
For our choice of kicking periods $\tau \equiv\hbar_{\rm eff} > 1$ \cite{I1990,F1993}, classical trajectories wandering about hierarchical structures of the classical phase space will not have a quantum analogue because those structures are too small to be resolved by the wave function.
This means that the observed fluctuations indeed arise from quantum localisation effects and not from a semiclassical diffusion process. 

\subsection{The kicking period as control parameter}
\label{sec:3.2}

As reviewed in section \ref{sec:2}, the sufficient conditions for the occurrence of fractal fluctuations are fulfilled for choices of $\tau$ for which the \dk{} exhibits dynamically localised behaviour. 
However, besides dynamical localisation the quantum \dk{} supports ``quantum resonant'' motion for specific values of $\tau$ and quasimomentum $\beta$ \cite{I1990,IS1980}. 
Our goal is to avoid as much as possible the impact of the quantum resonances on the dynamics, such that we can clearly identify the origin of the fractality of the survival probabilities. 
Since the parameter we scanned is the kicking period $\tau$, we verified that no signatures of quantum resonances are found in the analysed small range of $\tau$ and for the applied, finite kick numbers.

The quantum kicked rotor shows ballistic growth of momentum, shortly \emph{a quantum resonance}, if
\begin{eqnarray}
\label{eq:resonances}
\tau \in \{4\pi s/q ;\; s,q \in \mathbb{N}\}, 
\;\;\;\;\;\;\;  
\beta \in \{m/2s ,\; 0\leq m < s ;\; m,s \in \mathbb{N}\},
\end{eqnarray}
and in these cases the time dependence of energy on the number of kicks is \cite{I1990,IS1980}
\begin{eqnarray}
\label{eq:estimate}
E(t;\tau) = \eta t^2 + \mathcal{O}(t), 
\;\;\;\;\; \mbox{with} \;\;\;\;\;
\eta \simeq (k/q)^{2q}.
\end{eqnarray}
The denominator $q$ in the rational factor of $\tau$ is called the \emph{order} of the resonance. 
The set containing all the resonances has zero Lebesgue measure in any interval of kicking periods, but we do care about it because the dependence on $\tau$ of the survival probability $P_{\rm surv}(\tau;t)$ is continuous for a fixed, finite number of kicks and the fluctuations we want to observe should be caused by dynamical localisation and not by quantum resonances.
The preceding growth estimate (\ref{eq:estimate}) establishes that a resonance is suppressed for a time that increases more than exponentially with its order.
One way to avoid contributions from the resonances is to use a judicious choice of the range of $\tau$ and sampling grid $G$ used for
numerical simulations or experiments.
We chose
\begin{eqnarray}
G=\{\tau_{i}=\tau_{0}+i\cdot\delta\tau, i\in\{0,\dots,m-1\} \}, 
\;\;\;\;\; \mbox{with} \;\;\;\;\;  
m = 10^4 \;, 
\label{eq:grid}
\end{eqnarray}
where the value of $\tau_{0}\,/\,4\pi$ is a fraction of the golden mean:
\begin{eqnarray}
\label{eq:irrational}
\hspace{-20mm}
\frac{\tau_{0}}{4\pi} = \frac{14}{10}\frac{s}{q}(\sqrt{5}-1),
\;\;\;\;\;\;\;
s = 6142, \;\;\; q = 95403,
\;\;\;\;\;\;\;
\frac{s}{q} - \frac{1}{4\pi(\sqrt{5}-1)} < 10^{-11}\;.
\end{eqnarray}
For $\delta\tau = 9.98\times 10^{-7}$, also all other grid points in $G$ are incommensurable to
$4\pi$ up to the used significant digits.
We verified that the lowest order resonance in the 
range $[\tau_{\rm min},\tau_{\rm max}] \approx [1.4,1.41]$ has $q=107$ and that there is no crowding of resonances of order $q \leq 2000$ anywhere in this interval. 
Since we are not going to use times longer than $10^4$ kicks in our simulations, and kicking strength of order unity, the quadratic term is suppressed dramatically by the coefficient $\eta$ in (\ref{eq:estimate}), for all occurring resonances $q \geq 107 $.
Finally, we explicitly checked throughout the simulations that localisation is at work by inspecting the average energy and, for selected values of $\tau$, the shape of the wave function in momentum space, which shows a characteristic exponential decrease as explained below in section \ref{sec:3.3}.

We also tried a quantitative approach for the choice of the grid along the $\tau$ axis.
If some resonance were important, any numerical selection method could detect it and 
prefer grids with points away from the quantum resonances.
Our method is based on the maximisation in the ``grids space'' of a function $F(G(\tau_{0},\delta\tau))$ that adds a contribution from each resonance, up to a maximum order, within a given interval, and this contribution is the larger the farther the resonance position in $\tau$ [see Eq.~(\ref{eq:resonances})]
is from the nearest point of the grid.
This means that a ``higher mark'' is achieved by the grids whose points are away from the resonances.
Formally we defined
\begin{eqnarray}
S = \{4\pi s/q \}  \cap \{ q \leq q_{\rm max} \} \cap [\tau_{\rm min},\tau_{\rm max}] \nonumber \\
F(G(\tau_{0}, \delta\tau)) = \sum_{ \tau_{r} \in S } f_{r} (\min \{|\tau_{r}-\tau_{g}|;\tau_{g} \in G \}), \nonumber \\
f_{r}^{(0)}(\Delta\tau) = \Delta\tau ; \;\;\;\;\;
f_{r}^{(1)}(\Delta\tau) = {\Delta\tau}^2; \;\;\;\;\;
f_{r}^{(3)}(\Delta\tau) = \Delta\tau / q_{r}\;. \nonumber
\end{eqnarray}
Different definitions of the weight function $f_{r}(\Delta\tau)$ allow us to give more
weight to resonances with smaller $q \gtrsim 107$ (i.e., to those which influence the time evolution 
of a wider neighbourhood along the $\tau$ axis).
Of course, this programme requires detailed knowledge of the dynamics near the high-order resonances of $q \geq 107 $, but this goal has not been theoretically accomplished yet.
None of our weight functions could resolve the presence of a resonance by a sharp minimum when applied to a specific grid.

As a consequence of our choice of the interval of kicking periods and grid points in this interval, 
no signatures of quantum resonances are expected to manifest for interaction times of up to $10^4$ kicks.

\subsection{The opening of the system}
\label{sec:3.3}

The probability decay arises from the open geometry of our system, which is implemented mathematically by imposing absorbing boundary conditions in momentum space \cite{BCGT2001}. 
This means that 
$$\psi(n) \equiv 0 \;\;\;\;\;\mbox{if}\;\;\;\;\; n\leq n_{1}<0 \;\;\mbox{or}\;\;n\geq n_{2}>0.$$
The requirement on the boundaries is that they must guarantee dynamical localisation (see section \ref{sec:2}).
This happens if the wave function on the boundaries is ``so'' small that the kicking potential cannot spread a ``substantial'' part of the wave function out of the boundaries. 
The compatibility of the values of the parameters involved -- $t,k,n_{1},n_{2}$ -- is checked using a consequence of the conditions that grant a fractional dimension of the graph of the survival probability (see section \ref{sec:2}).
This consequence is that the square of the wave function decreases with time 
\emph{keeping its shape constant}, in the limited momentum lattice representing the open system.

Let us recall that the typical shape of a one-dimensional localised wave function is exponential, extending in a region intermediate between the support of the initial state in momentum space and the absorbing 
boundary.
In a linear-logarithmic plot the wave function is (apart from erratic fluctuations around its mean decrease) a line in this intermediate region; constancy of the shape means constancy of the steepness of the line.
This criterion, which is in fact a localisation criterion, was used as a 
prerequisite for all our simulations.
If the boundaries are too far away from the initial state, the decay is extremely slow (a consequence
of strong dynamical localisation).
To avoid long waiting times (which are hard to reach experimentally), asymmetric boundaries have been used, 
with $1 \approx |n_{1}| \ll |n_{2}|$, 
and a statistical initial ensemble of orbits at $t=0$ with $p=0$ and randomly distributed phases 
$\theta$, i.e., $\psi(n;t=0) \equiv \delta_{n,0}$. 
The wave function in momentum space $\psi(n;t)$ then evolves to a shape which is asymmetric with respect to $n=0$.
On the side where the cutoff is closer to the origin, the wave function does not decrease exponentially, 
and in a linear-logarithmic plot the momentum distribution shows a broad and smooth maximum, while at $n=0$ a sharp peak would be present if we choose $1 \gg |n_{1}| \approx |n_{2}| $.
Although the exponential decrease on the side of the larger cutoff $n_{2}$ is influenced by the opening at 
$n_{1}$, the shape indeed remains constant for a sufficiently large number of kicks in a 
range $[\bar{n},n_{2}[$, where the precise value of $\bar{n}\approx 50 \ldots 100$ 
depends on the choice of $n_{1}$.

\begin{figure} 
  \centering 
  \includegraphics[height=10cm]{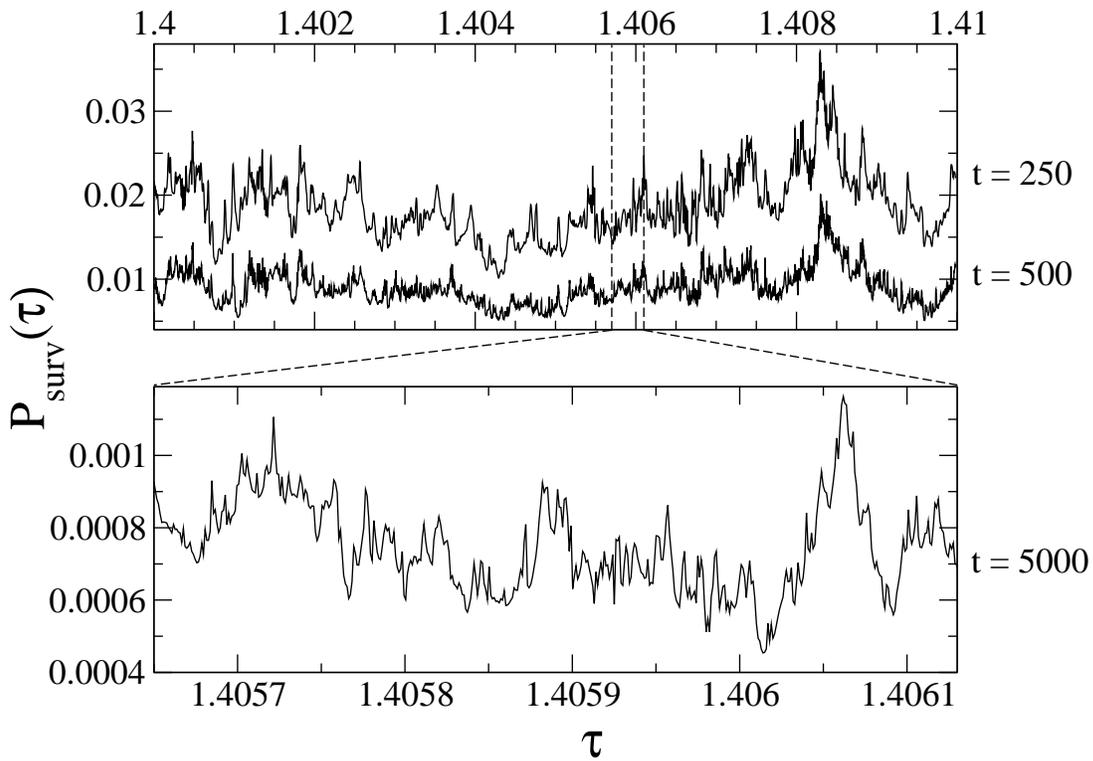} 
  \caption{
    The survival probability as a function of $\tau$ for $k=5$, $\beta=0$, $n_{1}=-1$, $n_{2}=200$ and different kick numbers $t$.
    The magnification in the lower panel shows that, as $t$ increases, self-affine fluctuations occur on finer and finer scales in $\tau$.
  }
  \label{fig:1} 
\end{figure}

\begin{figure} 
  \centering 
  \includegraphics[height=10cm]{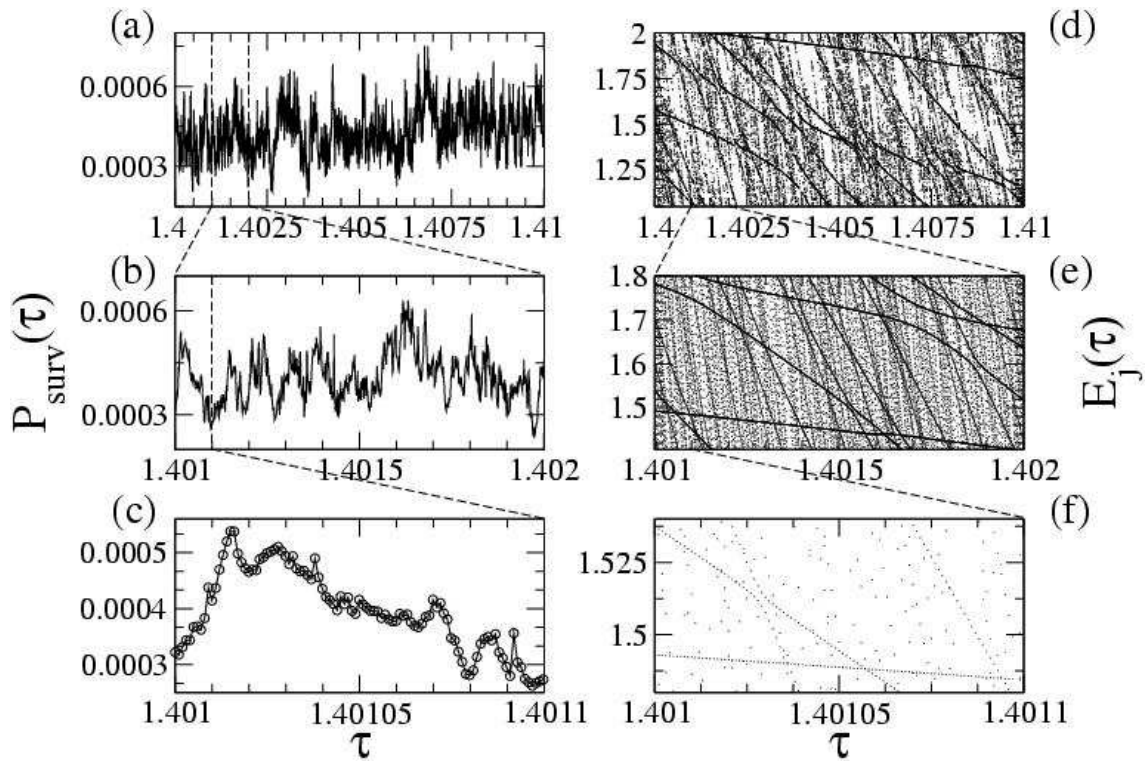} 
  \caption{
    (a,b,c) show the survival probability of figure \ref{fig:1} after $t=10^{4}$ kicks at different magnifications. For the same parameters, (d,e,f) show the real parts of the quasienergies as a function of $\tau$ (obtained as the eigenphases of the evolution operator (\ref{eq:evol}), which was represented in the basis of momentum states as a finite matrix in the range $n\in]n_{1},n_{2}[$ and then diagonalised). 
    We see that the fluctuations on finer and finer scales are accompanied by ubiquitous avoided-crossings in the eigenvalue spectrum (note that for better visibility in (d-f) only a small part of the full spectral range $[-\pi,\pi]$ is shown).  
  }
  \label{fig:2} 
\end{figure} 

\begin{figure} 
  \centering 
  \includegraphics[height=10cm]{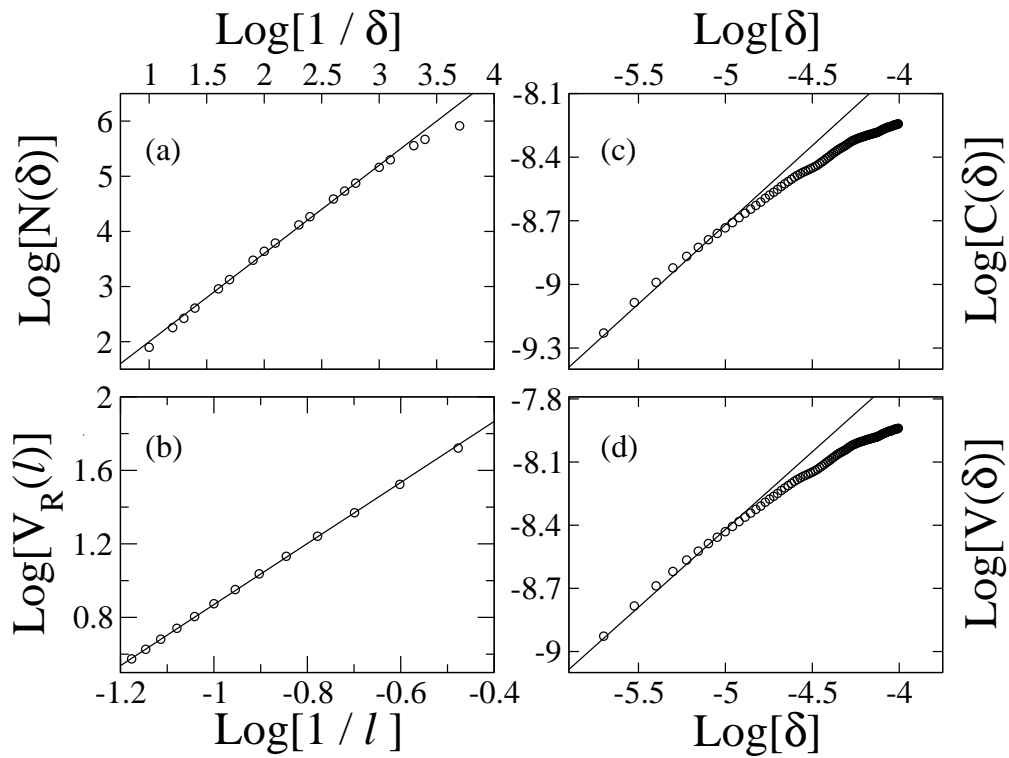} 
  \caption{
    Fractal analysis of the survival probability from figure \ref{fig:2} (a) using the following methods: (a) box-counting, (b) variational method, (c) correlations, (d) variances.
    The exponents of the fits (solid lines) are $D_{\rm f}= 1.6$ (a) and 1.7 (b), $a_{\rm corr}=0.8$ (c) and
    $a_{\rm var} = 0.8$ (d).
  }
  \label{fig:3} 
\end{figure}

\section{Numerical Results for fixed quasimomentum}
\label{sec:4}

The central result of this paper is the computation and fractal analysis of the survival probability $P_{\rm surv}(\tau; t, \beta, k, n_{1}, n_{2})$ as a function of $\tau$, while the other parameters are fixed for each curve. 
Our fractal analysis comprehends the computation of (a) the box-counting dimension \cite{GT2001,DQRT1989}, (b) a variational algorithm dimension \cite{DQRT1989}, together with the calculation of the (c) correlations, and (d) variances of the graph $P_{\rm surv}(\tau)$. 
Several curves are computed with different choices of parameters. 
Our results are essentially independent of quasimomentum $\beta$ and the applied boundaries $n_{1}$ and $n_{2}$, whose choice is guided by the considerations stated in section \ref{sec:3.3}.

Numerical algorithms, of course, do not distinguish the origin of the irregular profile of a fractal graph.
To make sure that the observed fractality is actually produced by quantum effects, we verified that the increase of $k$ in the range $2\ldots 6$ (for $\tau \gtrsim 1.4$) is accompanied by a monotonic 
increase of the fractal dimension.
This is a \emph{signature} of fractality owing to dynamical localisation of a weakly open quantum system.
As $k$ reaches a certain saturation value $k_{\rm sat}\approx 4.5$ (where $k\,\tau > 5$ exceeds the global chaos border \cite{I1990,F1993} and quantum chaos is fully developed) we verified that the fractal dimension ceases its substantial growth observed in the range $k=2\dots 4.5$.

At fixed kick number $t$, the survival probability is in principle a smooth function of $\tau$ on a sufficiently small scale $\delta\tau$.
After a grid in $\tau$ is chosen, fractality is expected to increase as $t$ increases due to the appearance of fluctuations on finer and finer scales.
A finer grid requires a longer time to yield a ``fractal graph'' down to finer scales, because it takes longer for the fluctuations to appear on a scale smaller than the grid resolution.
This scenario, where fractality is generated by ``dynamical intrusion'', is exemplified in figure \ref{fig:1} where the survival probability in the localised regime is shown after various interaction times.
The calculation of the fractal dimension as a function of time shows a monotonic increase from unity up to a value between $1.6$ and $1.7$.

We computed the survival probability $P_{\rm surv} (\tau)$ for various interaction times of 
up to $10^{4}$ kicks. The latter value is much larger than the kick numbers of the order 100 typically
realised in state-of-the-art experiments \cite{KOSR1999,SAGS2003}. 
Nevertheless, the monotonic behaviour in time can itself be used 
as an important signature of fractality.
In figure \ref{fig:2} (a-c) the profile of $P_{\rm surv}(\tau; t)$ is shown along with a small, yet representative part of three successive magnifications over two orders of magnitude in the kicking period $\tau$.
The real parts of the quasienergy spectrum are presented in figure \ref{fig:2} (d-f) in the same ranges of $\tau$.
The visible avoided crossings are a consequence of quantum chaotic dynamics and their ubiquitous presence on different scales in $\tau$ naturally compares to the self-affine fluctuations of the survival probability.
This comparison highlights the fact that the observed fractality is indeed a consequence of quantum chaos.

The box-counting plot in figure \ref{fig:3} (a) shows the number of adjacent squares $N(\delta)$ of width $\delta$ along the $\tau$ axis necessary to box all points of the curve from figure \ref{fig:2} (a). 
The scaling law $N(\delta) \sim \delta ^{-D_{\rm f}}$ thus determines the fractal dimension $D_{\rm f}$. 
The variational method (b) is a substantial refinement of the box-counting which typically gives more reliable results \cite{DQRT1989}.
It involves the division of the full analysed $\tau$ interval in $R$ subintervals, and the total variation of the curve on groups of $2l$ adjacent subintervals is computed. The average of these quantities is called $V_{R}(l)$ and the value of $R$ which gives the best scaling of the form $V_{R}(l) \sim l^{-D_{\rm f}}$ is used.
In addition to the direct fractal analysis of $P_{\rm surv}(\tau)$, we computed the autocorrelations and the variances of the fluctuating graphs.
The correlations $C(\Delta\tau) = \langle P_{\rm surv}(\tau)\cdot P_{\rm surv}(\tau+\Delta\tau) \rangle _{\tau}$ are shown in figure \ref{fig:3} (c), the variances $V(\Delta\tau)= \langle | P_{\rm surv}(\tau + \Delta\tau) - P_{\rm surv}(\tau)  |^{2} \rangle _{\tau}$ in figure \ref{fig:3} (d).
Recalling the power-law scaling of $\rho(\Gamma)$ (see section \ref{sec:2}), we can check the following set of relations:
\begin{eqnarray}
\label{eq:exponents}
\rho(\Gamma) \sim \Gamma ^{-\alpha} \;\;\;\;\;
\Rightarrow
D_{\rm f} \approx 1+\alpha\,/\,2 \;,
\end{eqnarray}
and
\begin{eqnarray}
C(\Delta\tau) - C(0) \sim {\Delta\tau}^{a}, \;\;\;\;\;
V(\Delta\tau) \sim {\Delta\tau}^{a} \;\;\;\mbox{with}\;\;\; D_{\rm f} = 2 - a\,/\,2 \;, \nonumber
\end{eqnarray}
in the presence of the numerically confirmed identity between the temporal decay exponent of $P_{\rm surv} (t) \propto t^{-a}$ and the exponent of the correlations \cite{BCGT2001,T2001}.
These relations can be used as alternative and independent routes to the determination of the fractal dimension $D_{\rm f}$.
This follows from the fractional Brownian motion nature of $P_{\rm surv}(\tau)$, which itself originates form the spectral properties of the opened \dk{} \cite{GT2001}, and which determines the $\Delta\tau\rightarrow 0$ properties of statistical quantities such as correlations and variances \cite{M1982}.

\begin{table}
\begin{center}
\begin{tabular}{|l|l|l|l|l|l|l|l|}
\hline
  $k$ & $\beta$ & $n_{1}$ & $n_{2}$ & $D_{\rm f,bc}$  & $D_{\rm f,v}$ & $a_{\rm corr}$ & $a_{\rm var}$ \nonumber  \\ \hline \hline
  2.0 &    0.0  &  -1     &   200   &   1.1    &   1.2   &     1.5    &    1.6    \nonumber  \\ \hline
  3.5 &    0.0  &  -1     &   200   &   1.3    &   1.4   &     1.0    &    1.0    \nonumber  \\ \hline
  4.0 &    0.0  &  -1     &   200   &   1.4    &   1.5   &     0.8    &    0.8    \nonumber  \\ \hline
  4.5 &    0.0  &  -1     &   200   &   1.5    &   1.6   &     0.8    &    0.8    \nonumber  \\ \hline
  5.0 &    0.0  &  -1     &   200   &   1.6    &   1.7   &     0.8    &    0.8    \nonumber  \\ \hline
  5.0 &  0.33   &  -1     &   200   &   1.6    &   1.7   &     0.8    &    0.8    \nonumber  \\ \hline
  5.0 &  0.38   &  -1     &   200   &   1.6    &   1.7   &     0.8    &    0.8    \nonumber  \\ \hline
  5.0 &    0.0  &  -1     &   250   &   1.6    &   1.7   &     0.8    &    0.7    \nonumber  \\ \hline
  5.0 &    0.0  &  -1     &   300   &   1.6    &   1.7   &     0.8    &    0.7    \nonumber  \\ \hline
  6.0 &    0.0  &  -1     &   300   &   1.6    &   1.7   &     0.8    &    0.7    \nonumber  \\ \hline
\end{tabular}
\caption{
  Fractal analysis of the survival probabilities after $10^4$ kicks. 
  $D_{\rm f,bc}$ states to box counting dimension, while $D_{\rm f,v}$ 
  is obtained via the variational method.
  $a_{\rm corr}$ and $a_{\rm var}$ are the exponents of the fits to the correlations and variances, respectively.
  The estimated uncertainty derived from our fits is $\pm 0.1$ for the fractal dimensions as well as the exponents.
  } 
\end{center}
\end{table}

Table 1 reports the fractal dimensions which were obtained by the above four methods. $D_{\rm f,bc}$ 
and $D_{\rm f,v}$ are the box-counting and the variational dimension, respectively, while the exponents of the  correlations and variances are denoted $a_{\rm corr}$ and $a_{\rm var}$.
The table highlights the features already mentioned, i.e., the increase of $D_{\rm f}$ for increasing kicking strength $k$ and its basic independence of quasimomentum and the chosen cutoffs.
The fractal dimension saturates for $k\geq k_{\rm sat} \gtrapprox 4.5$.
We verified this saturation with a series of simulations conducted for $13$ values of $k \in [2,6]$
(not all shown in table 1).

The obtained four independent methods of our fractal analysis (summarised in table 1) give fairly consistent results with each other, with an estimated precision of $\pm 0.1$.
A systematical underestimation by box-counting method is observed, but also expected \cite{DQRT1989}
when applying it to curves with $D\gtrapprox 1.5$.

For our choice of the grid in $\tau$ [see Eq.~(\ref{eq:grid})] we noticed by inspecting the correlations and variances that, for $k\gtrapprox 5$, not all the fluctuations of the true curve are resolved by our grid.
This yielded systematically smaller and meaningless values for $a$, a problem which does not affect the box-counting and variational method that do not depend so critically upon the values of neighbouring points of the analysed graph.
Augmenting the resolution of our grid in $\tau$ on a test interval $[\tau_{0}, \tau_{0} + 10^{3}\,\delta\tau]$ (c.f. Eq.~(\ref{eq:irrational}) for the definition of $\tau_{0}$ and $\delta\tau$) we nevertheless were able to estimate the exponents of the correlations and the variances for $k\geq 5$ and $n_{2}\geq 250$ shown in table 1.

As a final test of our hypothesis that no trace of quantum resonances can be observed for the chosen interval in $\tau$ and our maximal interaction time of $10^{4}$ kicks, we analysed the survival probability for $k=5$ for two different quasimomenta $\beta \approx 1/3$ and $\beta=0.378942469767714$ 
(stated as $0.33$ and $0.38$, respectively, in table 1).
The latter value was chosen as a fraction of the golden mean 
to avoid any resonance condition in $\beta$ [see Eq.~(\ref{eq:resonances})].
As can be seen from table 1, no dependence on quasimomentum is found for the 
dynamically localised regime ($k=5$). 

In this section we presented a full-featured analysis of the fractal dimension of the survival probability 
$P_{\rm surv}(\tau)$, studied the dependence on the parameters $t$ and $k$ and observed how these dependences provide systematical signatures of fractality caused by quantum effects.
We found that $P_{\rm surv}(\tau)$ is indeed fractal over a substantial range of scales, and its dimension can be estimated between $1.6$ and $1.7$.
These numbers are stable when varying the initial quasimomentum (which is a constant of the motion) and the 
selected locations of the cutoffs $n_{1}$ and $n_{2}$.
Having in mind that the numerical determination of the fractal dimension of a graph bears some finite error, our results are consistent with the fractal dimension $1.5$ found for fixed $\tau=1.4$ 
in the scan of quasimomentum \cite{BCGT2001}.
The tendency towards a slightly larger fractal dimension in our data could be related to the distribution of decay widths, whose precise form is sensitive to the chosen values of $n_{1}$ and $n_{2}$ (see 
Ref. \cite{WKB2002} and discussion in section \ref{sec:2}).

\begin{figure} 
  \centering 
  \includegraphics[height=10cm]{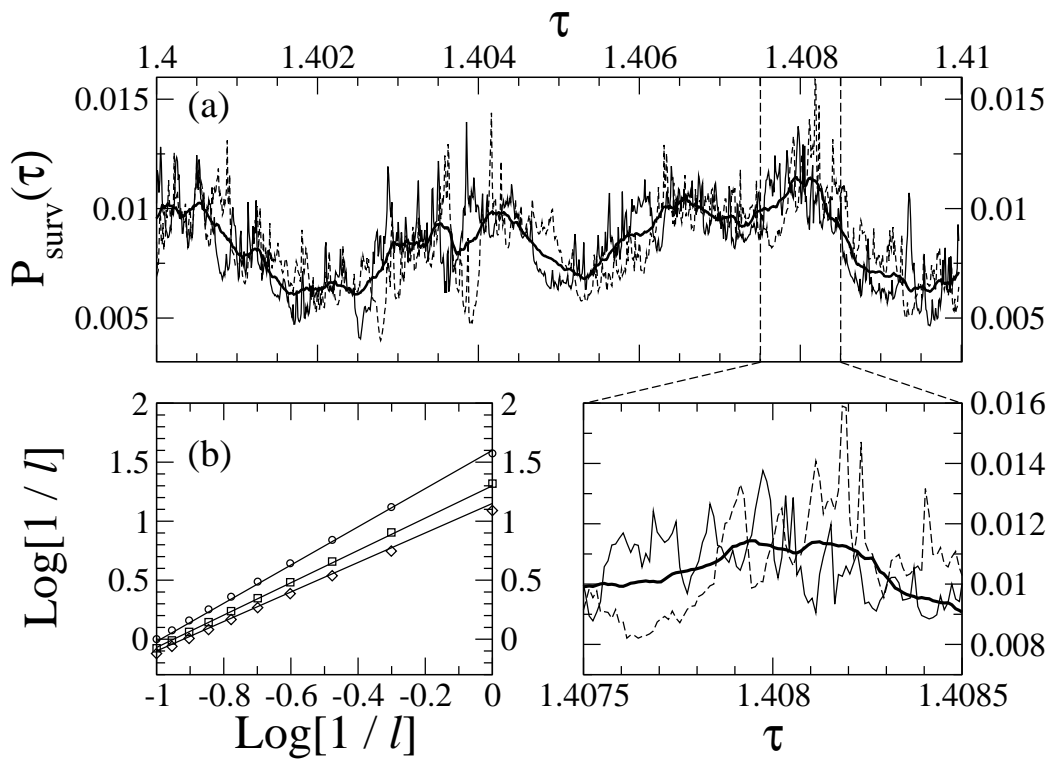} 
  \caption{
    (a) shows two survival probabilities for fixed $\beta\approx 0.006554$ and $\beta \approx 0.002009$ together with an average of $10^3$ $\beta$ values equally distributed in $[0,0.01]$ at $t=500$, $k=4.5$, $n_{1}=-1$, $n_{2}=200$.
    The average curve (thick) is smoother but its fractional dimension is nevertheless greater than unity.
    (b) shows the fractal analysis by the variational method for $\beta\approx 0.006554$ (circles) 
     that yields $D_{\rm f} \approx 1.6$, for the average of $10^3$ $\beta\in[0,0.01]$ (diamonds) with
     $D_{\rm f} \approx 1.2$, and for the average of only $10$ values of $\beta$ in the same interval 
     (squares) that gives the intermediate value of $D_{\rm f} \approx 1.4$.}
  \label{fig:4} 
\end{figure}

\begin{figure} 
  \centering 
  \includegraphics[height=10cm]{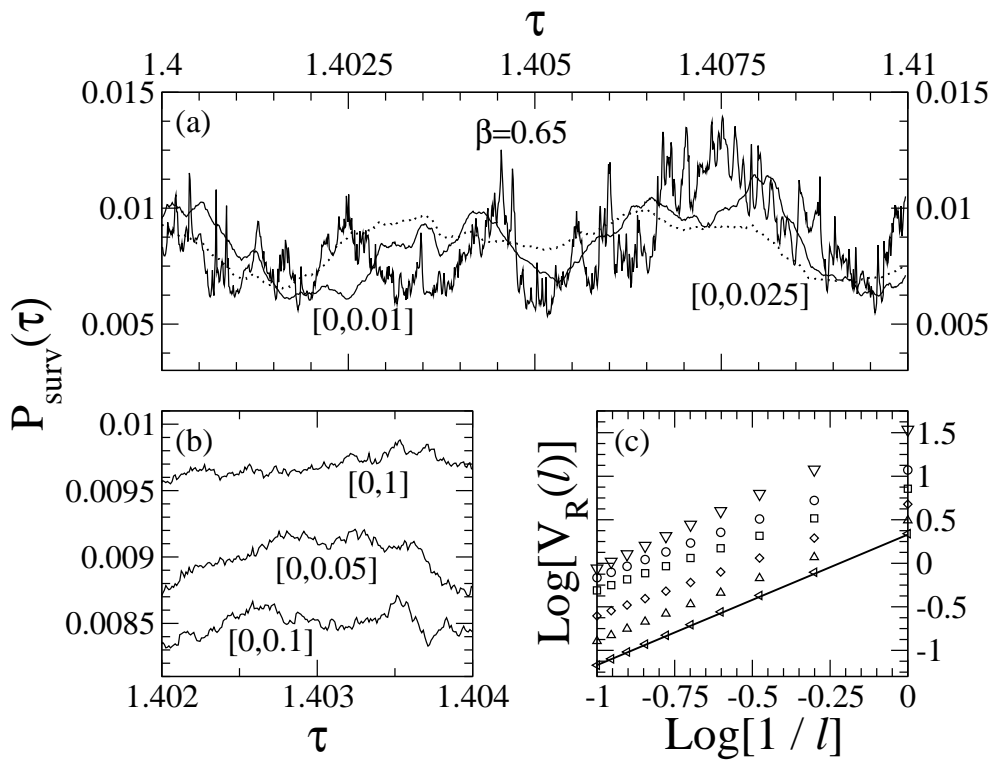} 
  \caption{
    (a) average of the survival probability (for $k=4.5$ and after $500$ kicks) over $10^{3}$ values of $\beta$ uniformly distributed in $[0,0.01]$ (solid) and $[0,0.025]$ (dotted) together with a survival probability for a fixed value $\beta\approx0.65$. 
    (b) Same as (a) for $10^{3}$ values of $\beta$ uniformly distributed in the shown intervals.
    (c) Fractal analysis by the variational method for the survival probabilities shown in (a,b). The fractal dimensions  are obtained by linear fits (shown only for $\Delta \beta =1$) through the symbols 
 $D_{\rm f}\approx 1.6$ (inverse pyramids, $\beta \approx 0.65$), 
$1.2$ (circles, $\beta \in [0,0.01]$), $1.2$ (squares, $\beta \in [0,0.025]$), $1.3$ 
(diamonds, $\beta \in [0,0.05]$), $1.4$ (pyramids, $\beta \in [0,0.1]$), and $1.5$ 
(left triangles, $\beta \in [0,1]$).
  }
  \label{fig:5} 
\end{figure} 

\begin{figure} 
  \centering 
  \includegraphics[height=10cm]{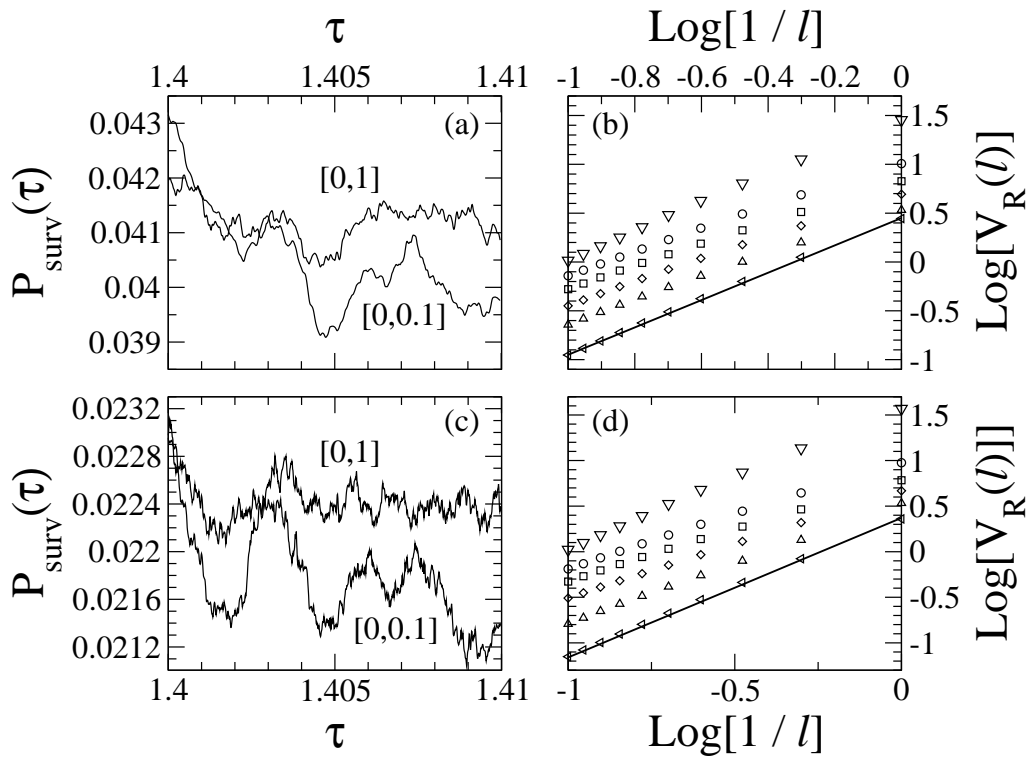} 
  \caption{ 
    (a,c) average of the survival probability over $10^{3}$ values of $\beta$ uniformly distributed in $[0,0.1]$ and $[0,1]$, after (a) $100$ and (c) $200$ kicks and for $k=4.5$.
    The enhancement of self-affine fluctuations with time is clearly visible.
    (b) and (d) show the fractal analysis by the variational method for (a) and (b), respectively,
    corresponding to $\beta\approx 0.65$ (inverse pyramids), or $10^{3}$ values of $\beta$ uniformly distributed in $[0.01]$ (circles), $[0.025]$ (squares), $[0,0.05]$ (diamonds), $[0,0.1]$ (pyramids), $[0,1]$ (left triangles).
    The fractal dimensions are $D_{\rm f} \approx 1.5$, $1.2$, $1.1$, $1.2$, $1.2$, $1.4$ in (b) and
    $1.6$, $1.2$, $1.1$, $1.2$, $1.3$, $1.5$ in (d) for increasing 
    width of the $\beta$ distribution. 
  }
  \label{fig:6} 
\end{figure}

\section{Signatures of fractality for realistic experimental conditions}
\label{sec:5}

\subsection{Experimental control of parameters}
\label{sec:5.1}

To realise an experiment where the fractal dimension of the survival probability, as studied in the preceding section, can be measured, it is necessary to address some principal problems of atom-optics kicked rotor experiments.

Control over the kicking strength $k$ is granted with a precision of a few percent \cite{BRMS1999,AGOS2001}.
Anyway, table 1 tells us that a variation of $k$ of the order up to $25\%$ is not crucial.
Time is one of the best controlled experimental parameters, and this feature makes it an ideal candidate for implementing an experiment to search for fractal fluctuations.
Kicking periods between about hundred nanoseconds and a few hundred microseconds are available, 
with a maximal precision of a few nanoseconds \cite{SWPL2005,AGOS2001,OSR2003,RAVA2005}. For caesium atoms, 
this range corresponds to dimensionless kicking periods (see section \ref{sec:1}) 
$\tau \approx 10^{-2} \ldots 18 $, and
a maximal precision of $\delta \tau \gtrsim 10^{-4}$. This precision implies that about 100
points could be scanned in our analysed interval in $\tau$, which would be sufficient for a rough,
qualitative verification of our predictions.

Any experiment will have a finite detection window of observable momentum classes.
The actual width of this window is typically determined by the imaging resolution and by the minimal signal-to-noise ratio of the measurement device \cite{KOSR1999,AGOS2001,AGSG2004}.
The detection window also determines a maximum interaction time after which the detection of a constantly decreasing atomic ensemble (due to the open boundary conditions) becomes meaningless.
In other words, the maximum number of kicks is limited by the precision disposable in the determination of the final momentum distribution.
Correspondingly, in our results reported below we choose the minimal kicking strength $k=4.5$ where the fractal dimension starts to saturate (see section \ref{sec:4}) and a maximum interaction time of $500$ kicks. The latter implies that we can choose a wider grid in $\tau$ because very fine structures do not develop for interaction times $t\leq 500$.
We used $\delta\tau^{\prime}=10\;\delta\tau$, $\tau_{0}^{\prime} = 
\tau_{0}$ and $m^{\prime} = m\, /\, 10$ [c.f. Eq.~(\ref{eq:irrational})].
A problem will certainly be the realisation of our idealised absorbing boundary conditions at specific momentum classes of the atoms.
Here methods using, for instance, external cutting potentials -- such as so called radio-frequency knives \cite{SHBC2005} or equally operating additional lasers -- could be thought of.

\subsection{The experimental initial ensemble}
\label{sec:5.2}

To approach real experimental scenarios, we shall analyse the survival probability for a smaller number of kicks of order $100$ \cite{KOSR1999,SAGS2003} and take into account an initial spread of quasimomentum among the ensemble of cold atoms \cite{WGF2003,AGSG2004}.

For a typical ensemble of cold atoms, the momentum distribution is Gaussian-like, with a width exceeding that of the Brillouin zone $2\,\hbar k_{L}$, equal to $1$ in our dimensionless units \cite{KOSR1999,SWPL2005,BRMS1999,AGOS2001,AGSG2004,WSPL2005,SAGS2003}.
Folding produces approximately a uniform distribution in the entire Brillouin zone, i.e., a uniform distribution of quasimomenta with a width of $\Delta\beta=1$ \cite{WGF2003}.
Using atoms in the Bose-Einstein condensate phase as initial ensemble allows the experimentalist a much better control over the width of the quasimomentum distribution \cite{WMMA2005}.
Values of $\Delta\beta \lessapprox 0.05$ have been realised in this context \cite{RAVA2005,DMCW2004,DPMS2004}.
Letting the condensate expand a little before the actual kicking evolution, allows one to reduce the atom-atom interactions to negligible values, with only slight changes in $\Delta\beta$ \cite{DPMS2004}.
As a consequence, the survival probability, experimentally measured by counting the number of atoms contained within the finite detection window, would be the result of an average of many \emph{independent} survival probabilities with different values of quasimomentum.
The independence of probabilities follows from the independent dynamics of the atoms  \cite{KOSR1999,SWPL2005,BRMS1999,AGOS2001,AGSG2004,WSPL2005,DPMS2004}, while the coherent evolution of a single atoms is still essential for the observed behaviour.

We computed $P_{\rm surv}(\tau; \beta)$ for different ranges of $\beta$, and then averaged the resulting curves to arrive at $\langle P_{\rm surv}(\tau; \beta) \rangle_{\beta}$.
Figure \ref{fig:4} investigates the effect of averaging over $\beta$ on the fractal dimension.
The survival probabilities for two fixed $\beta$ are shown, together with the average for a uniform distribution of $10^{3}$ values of $\beta\in[0,0.01]$.
Figure \ref{fig:4} (a) shows that the average curve is quite smooth on large scales, but nevertheless presents fluctuations on finer resolutions, with a fractal dimension substantially larger than unity.
We verified that, by decreasing the number of atoms in the ensemble, the dimension steadily increases.
We encounter a signature of fractality, which experiments could detect even far from the idealised limit of the one-atom dynamics.
That is, the fluctuating behaviours of the averaged curves is a direct consequence of fluctuations of single $\beta$ curves.

Figure \ref{fig:5} (a) and (b) show the average survival probability for ensembles with the same number of $\beta$ values but with different, larger widths $\Delta\beta$ of the initial quasimomentum distribution.
Wider distributions are smoother on large scales and are not drawn in figure \ref{fig:5} (a) because they could not be appreciated by eye when compared to curves for $\Delta\beta=0.010$ and $0.025$.
The magnification in (b) shows that the fluctuations exhibit smaller excursions.
A fractal analysis [see figure \ref{fig:5} (c)] by the variational method shows that the dimension $D_{\rm f}$ remains in all cases larger than unity and, moreover, does \emph{not} vary monotonically as $\Delta\beta$ is increased.

We interprete our results for finite $\Delta\beta$ in the following way: while a small range $\Delta\beta$ tends to wash out the fractal behaviour of the curves with one fixed $\beta$, an average over larger ranges $\Delta\beta$ tends to lift the fractal dimension again.
This line follows nicely from the prediction of \cite{GT2001} where it is argued that fractality can arise from superimposing non-fractal patterns on appropriate scales of the scanned variable \cite{T2001,GTWunp}.

Figure \ref{fig:6} repeats the analysis of figure \ref{fig:5} for $t=100$ (a,b) and $t=200$ (c,d).
The fractal dimension of each average survival probability for a definite value of $\Delta\beta$ is seen to be a monotonic function of time, what
points out once more the dynamical origin of the analysed fluctuations.
This contrasts the dependence of the fractal dimension on $\Delta\beta$ at fixed time, which is non monotonic, because averaging both washes out the fluctuations of the single curves for small $\Delta\beta\lessapprox 0.1$, while it creates new ones by superimposition for $\Delta\beta\gtrapprox 0.1$.

\begin{figure} 
  \centering 
  \includegraphics[height=6.5cm]{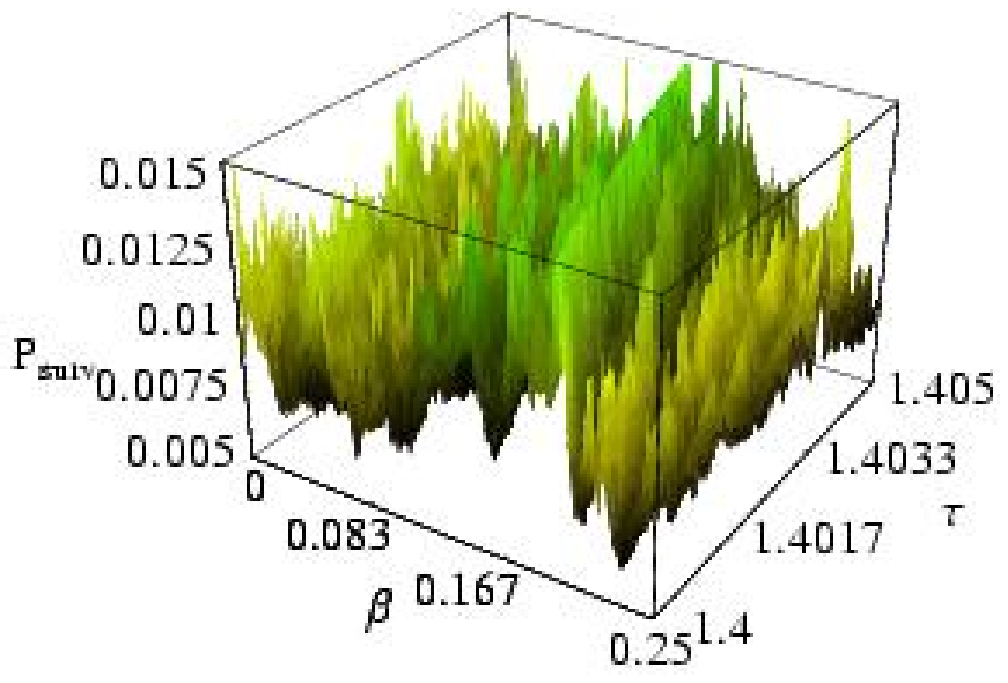} 
  \includegraphics[height=6.5cm]{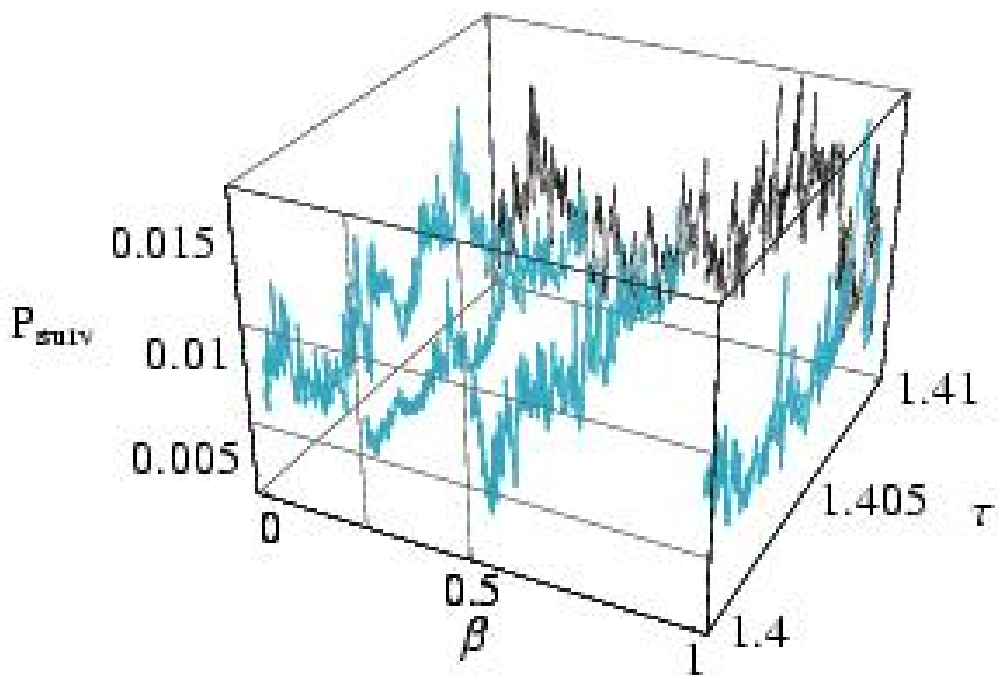}
  \caption{
    (color online).
    On the left panel the survival probability as a function of $\beta$ and $\tau$ 
    is shown after $500$ kicks, for $k=4.5$, $n_{1}=-1$ and $n_{2}=200$.
    Slices of the graph $P_{\rm surv}(\beta,\tau)$ are shown on the right panel.
    The thick lines represent the survival probability as a function of $\tau$ that is analysed in section \ref{sec:4} (we have used $10^{3}$ similar curves to compute the incoherent average which is the experimental observable, as explained in section \ref{sec:5.2}).
    The thin curve lying in the plane orthogonal to the $\tau$ axis is the survival probability $P_{\rm surv}(\beta)$ as a function of the quasimomentum $\beta$ such as studied in \cite{BCGT2001}.
  }
  \label{fig:7} 
\end{figure}

From a more general perspective, the survival probability $P_{\rm surv}(\beta,\tau)$ can be seen as a surface lying over the plane spanned by the two variables $\tau$ and $\beta$.
In \cite{BCGT2001} $P_{\rm surv}(\beta,\tau)$ was analysed at fixed $\tau$ using $\beta$ as scanning parameter, i.e., a slice of the surface parallel to the $\beta$ axis was analysed.
In section \ref{sec:4} we studied the ``orthogonal'' problem of fixed $\beta$, using $\tau$ as a scanning variable.
Averaging over $\beta$ can be interpreted as a ``column density'', i.e., $P_{\rm surv}(\beta,\tau)$ integrated over the $\beta$ degree of freedom.
The resulting averaged curve is the observable experimentally accessible as discussed in this section.
Looking at figure \ref{fig:7}, our results can thus be interpreted geometrically: the fractal behaviour of the slices $P_{\rm surv}(\tau ; \beta = \mbox{const})$ is to a large extent preserved by the average over a typical experimental spread in $\beta$.

\section{Conclusions}
\label{sec:6}

We considered the quantum kicked rotor, a paradigmatic model of quantum chaos, which describes the time evolution of noninteracting cold atoms in periodically flashed optical lattices.
Imposing absorbing boundary conditions allows one to probe the transport properties of the system, and in particular to define the survival probability of atoms on a finite region in momentum space.
For fixed kick numbers, the quantum survival probability depends sensitively on the parameters of the system, and a self-affine structure of the survival probability $P_{\rm surv}$ is predicted, as either the kicking period or quasimomentum is scanned.

Instead of using the initial quasimomentum $\beta$ as control parameter, as done in the numerical simulations of \cite{BCGT2001}, we used the kicking period $\tau$ as scan parameter, which is much better controllable experimentally. 
We verified the fractal nature of the graph of the survival probability $P_{\rm surv}(\tau)$ in the dynamically localised regime, and obtained a fractal dimension $D_{\rm f}\approx 1.6 \pm 0.1$ for large but finite interaction times, for which quantum resonances do not manifest.

Any experimental setup prepares cold atoms with a finite spread in quasimomentum.
The experimental observable is then the average of the survival probabilities over the quasimomentum distribution.
We reproduced this observable by computing the incoherent average $\langle P_{\rm surv}(\tau) \rangle_{\beta}$, and found that the fractal dimension of the average remains substantially larger than unity, even for shorter interaction times of a few hundred kicks.

We conclude that the fractality in the survival probability induced by quantum chaos is an unexpectedly robust feature and in spite of many challenging aspects (see section \ref{sec:5}) could be observed in a future atom-optics experiment.
Apart from the experimental verification of fractal fluctuations of purely quantum origin, a remaining open problem is whether a universal scaling law for the fractal dimension could be found as a function of {\em both} parameters $\tau$ and $\beta$, including quantitative predictions for the here computed averages $\langle P_{\rm surv}(\tau,\beta) \rangle_{\beta}$ over a finite range of $\beta$.

\ack 
We thank Emil Persson for bringing the variational method for the fractality analysis (see \cite{DQRT1989}) to our attention.
S.W. acknowledges support by the Alexander von Humboldt Foundation (Feodor-Lynen Program) and is grateful to Italo Guarneri for a profound introduction to the field of ``quantum fractals''.

\end{document}